# Spin Splitting of Dopant Edge States in Magnetic Zigzag Graphene Nanoribbons

Raymond E. Blackwell[1,†], Fangzhou Zhao[2,3†], Erin Brooks[1], Junmian Zhu[1], Ilya Piskun[1], Shenkai Wang[1], Aidan Delgado[1], Yea-Lee Lee[2], Steven G. Louie[2,3,*] & Felix R. Fischer[1,3,4,*]

[1]Department of Chemistry, University of California, Berkeley, CA 94720, USA. [2]Department of Physics, University of California, Berkeley, CA 94720, USA. [3]Materials Sciences Division, Lawrence Berkeley National Laboratory, Berkeley, CA 94720, USA. [4]Kavli Energy NanoSciences Institute at the University of California Berkeley and the Lawrence Berkeley National Laboratory, Berkeley, California 94720, USA.

[†] These authors contributed equally to this work.

[*] Corresponding authors

Spin-ordered electronic states in hydrogen-terminated zigzag nanographene give rise to magnetic quantum phenomena[1,2] that have sparked renewed interest in carbon-based spintronics[3,4]. Zigzag graphene nanoribbons (ZGNRs) — quasi one-dimensional semiconducting strips of graphene featuring two parallel zigzag edges along the main axis of the ribbon — are predicted to host intrinsic electronic edge states that are ferromagnetically ordered along the edges of the ribbon and antiferromagnetically coupled across its width[1,2,5]. Despite recent advances in the bottom-up synthesis of atomically-precise ZGNRs, their unique electronic structure has thus far been obscured from direct observations by the innate chemical reactivity of spin-ordered edge states[6-11]. Here we present a general technique for passivating the chemically highly reactive spin-polarized edge states by introducing a superlattice of substitutional nitrogen-dopants along the edges of a ZGNR. First-principles GW calculations and scanning tunneling spectroscopy reveal a giant spin splitting of the low-lying nitrogen lone-pair flat bands by a large exchange field (~850 Tesla) induced by the spin-polarized ferromagnetically ordered edges of ZGNRs. Our findings directly corroborate the nature of the predicted emergent magnetic order in ZGNRs and provide a robust platform for their exploration and functional integration into nanoscale sensing and logic devices[11-17].

Graphene nanostructures terminated by zigzag edges host spin-ordered electronic states that give rise to quantum magnetism[1,2]. These intrinsic magnetic edge states emerge from the zigzag edge structure of graphene itself, and create opportunities for the exploration of carbon-based spintronics and qubits[18-20], paving the way for the realization of high-speed, low-power operation spin-logic devices for data storage and information processing[21-24]. The edge states of zigzag graphene nanoribbons (ZGNRs) have been predicted to exhibit a parallel (ferromagnetic) alignment of spins on either edge of the ribbon while the spins on opposing edges are antiferromagnetically coupled (antiparallel alignment)[1,2]. This unusual electronic structure can give rise to field- or strain-driven half-metallicity in ZGNRs[2,25]. A strong hybridization of the electronic states of ZGNRs with those of the underlying support, along with the susceptibility of zigzag edges to undergo passivation through atom-abstraction or radical-recombination reactions represents a veritable challenge to their exploration.

Our strategy for engineering chemically robust ZGNR edge states relies on the introduction of a superlattice of isoelectronic substitutional dopant atoms along both edges of a ZGNR. Replacement of every sixth C–H group along the zigzag edge of a 6-ZGNR (a zigzag GNR featuring six lines of carbon atoms across the width of the ribbon) by a N-atom leads to the structure of the N-6-ZGNR depicted in Figure 1a. The N-atoms disrupt the extended zigzag edge states to shorter segments reminiscent of the zigzag edge of pentacene — arguably one of the largest chemically persistent unsubstituted acenes. Each trigonal planar N-atom contributes the same number of electrons (one electron in a half-filled $p_z$-orbital) to the extended π-system of the N-6-ZGNR as the trigonal planar C–H groups they replace. We anticipate that the magnetic spin-polarized edge states remain largely unaffected while a subtle modulation of frontier bands imposed by a superlattice of N-dopant atoms, having lower energy $2p_z$ orbitals than carbon, leads to a passivation of the reactive zigzag edge.

Guided by this idea we designed a molecular precursor for N-6-ZGNRs, the dibenzoacridine **1** (Fig. 1b). N-6-ZGNRs were grown on Au(111)/mica films by sublimation of **1** in UHV onto a clean Au(111) surface held at 297 K. Figure 1c shows a constant-current scanning tunnelling microscopy (STM) topographic image of a sub-monolayer coverage of **1** on Au(111) at $T = 4$ K. Molecule-decorated surfaces were subsequently annealed at 475 K to induce the homolytic cleavage of C–Br bonds followed by the radical step-growth polymerization to give *poly*-**1**. Further annealing at 650 K induces a thermal cyclodehydrogenation that leads to the fully fused N-6-ZGNRs (Fig. 1d). STM topographic images reveal extended GNRs featuring atomically smooth zigzag edges with an apparent height and width of 0.23 nm ± 0.03 nm and 1.95 nm ± 0.05 nm, respectively, consistent with the formation of the fully conjugated N-6-ZGNR backbone (Fig. S1). In order to infer the position of the N-atoms along the edges of N-6-ZGNRs, we relied on the presence of a characteristic edge-defect, previously observed for all-carbon 6-ZGNRs[11]. The excision defect depicted in Figure 1e emerges from the homolytic cleavage of a *m*-xylene group in *poly*-**1** during the thermal cyclodehydrogenation step and results in a concave indentation along the zigzag edge of the GNR (see inset in Fig. 1e). Based on the chemical structure of the molecular building block **1** we can assign the location of a N-dopant atom to the opposing zigzag edge across from the defect site.

Topographic (Fig. 2a) and bond-resolved STM (BRSTM) images (Fig. 2b) using CO-functionalized STM tips were recorded on a fully cyclized segment of N-6-ZGNR featuring the same *m*-xylene deletion defect described above. While the topographic STM image (Fig. 2a) resolves the zigzag edge structure of N-6-ZGNRs and hints at a superlattice associated with the position of substitutional N-dopants, the BRSTM image (Fig. 2b) shows an alternating pattern of five bright lobes protruding from the edge of the N-6-ZGNRs (arrows in Fig. 2b) flanked on either side by indentations of darker contrast. Most notably the pattern on one zigzag edge is offset by ½ period from the opposite edge and is superimposable with the position of N-atoms derived from the analysis of *m*-xylene deletion defects. The enhanced signal in zero-bias d$I$/d$V$ imaging suggests a strong hybridization of the Au(111) surface state with the magnetic edge states at the bandgap of the ribbon. d$I$/d$V$ point spectra recorded along the edges of N-6-ZGNRs show only a broad featureless local density of states (LDOS) that cannot be assigned to the van Hove singularities of the valence band (VB) and conduction band (CB) edge states (Fig. 3a, Fig. S2b).

To overcome this robust electronic coupling and to experimentally access the magnetic edge states of an isolated ZGNR, we developed a STM tip-induced decoupling protocol that irreversibly disrupts the strong hybridization of the N-6-ZGNRs with the Au(111) surface. When placing an STM tip ~ 4 Å above the centre of the N-6-ZGNR (red cross in Fig. 2a), and sweeping the bias voltage from $V_s = 0.00$ V to $V_s = +2.50$ V, a discontinuous drop ($\Delta I_t^+ = 0.16$ nA) in the tunnelling current can be observed at a bias of $V_s^+ = +2.23 \pm 0.05$ V (Fig. 2e, Fig. S3). The abrupt shift in the tunnelling current suggests an electronic decoupling of the GNR from the Au(111) substrate. Subsequent bias sweeps from $V_s = 0.00$ V to $+2.50$ V near the position of the red cross yield no further change in the tunnelling current. STM imaging of the same N-6-ZGNR segment following SPM tip-induced decoupling reveals a local change in the constant-height d$I$/d$V$ map (Fig. S4). In the BRSTM image (Fig. 2c), the immediate area of the ribbon in the vicinity of the STM tip during the decoupling step (1–2 nm surrounding the position of the red cross) is clearly resolved and shows the distinctive structure of the N-6-ZGNR backbone as well as the characteristic pattern of N-atoms along both zigzag edges (arrows in Fig. 2c; N-atoms appear with a darker contrast when compared to the C–H groups; Fig. S5). The same irreversible decoupling event can be observed by applying a negative bias. Sweeping the bias voltage from $V_s = 0.00$ V to $V_s = -2.50$ V at the position marked by a blue cross in Figure 2a, reveals a comparable drop in the tunnelling current at $V_s^- = -2.16 \pm 0.05$ V ($\Delta I_t^- = 0.15$ nA). The resulting BRSTM image (Fig. 2d) shows the bond-resolved structure of the N-6-ZGNR backbone along the entire length of the ribbon. Neighboring ribbons that are not covalently fused with the N-6-ZGNR subjected to the decoupling protocol remain unaffected by this process (Fig. S6).

The local electronic structure of surface-decoupled N-6-ZGNRs was characterized using d$I$/d$V$ point spectroscopy. A typical d$I$/d$V$ point spectrum recorded along the edge of a N-6-ZGNRs (> 3 nm from either end of the ribbon; see inset in Fig. 3a) shows two prominent electronic states, a sharp peak centred at $V_s = -0.30 \pm 0.02$ V and a broader feature centred at $V_s = +0.50 \pm 0.05$ V. Comparing to our GW calculation results, the peaks at $V_s = -0.30$ V and $+0.50$ V can be identified as the N-6-ZGNR VB and CB edge, respectively. The resulting bandgap for a nitrogen edge-doped N-6-ZGNRs is $\Delta E_{exp} = 0.80 \pm 0.05$ eV, ~0.7 eV smaller than the experimental bandgap of a pristine all-carbon 6-ZGNRs ($\Delta E_{exp} = 1.5$ eV)[11]. d$I$/d$V$ imaging of the spatial distribution of the N-6-ZGNR LDOS at energies close to the CB edge (Fig. 3b) shows the largest contrast along the zigzag edge C–H groups immediately flanking the position of N-atom dopants while the intensity decreases toward the centre of the N-6-ZGNR backbone. d$I$/d$V$ maps recorded at an imaging bias of –0.30 V reveal the LDOS associated with the VB state (Fig. 3b) evenly distributed over the C–H groups lining the zigzag edges. Both d$I$/d$V$ maps of the CB and VB edges show a weaker contrast at the position of the N-atom dopants further corroborating the successful segmentation of the magnetic edge state into a superlattice of pentacene-like fragments.

Experimental results are in excellent agreement with theoretical calculations based on *ab initio* density functional theory (DFT) within the local spin density approximation (LSDA),[26] and *ab initio* GW calculations[27] which includes the important self-energy corrections to the quasiparticle excitations measured in STS experiments. The first-principles results provide quantitative evidence that the SPM tip-induced decoupling has resulted in a full recovery of the intrinsic magnetic edge states (both in energy and wavefunction) of N-6-ZGNRs. Figures 3c,e show the calculated LDOS maps at a distance of 4 Å above the plane of the freestanding N-6-ZGNR at energies corresponding to the CB and VB edges. The characteristic pattern and relative contrast of protrusions lining the zigzag edges of N-6-ZGNRs seen in the experimental d$I$/d$V$ maps of CB and VB states (Figs. 3b,d) are faithfully reproduced in the corresponding GW LDOS maps (Figs. 3c,e). The GW band structure and the corresponding density of states (DOS) are depicted in Figures 4a and 3f, respectively. The quasiparticle bandgap $\Delta E_{GW}$ = 0.83 eV agrees well with the experimental gap $\Delta E_{exp}$ = 0.80 ± 0.05 eV derived from STS. The agreement between experimental d$I$/d$V$ maps and the theoretically predicted LDOS, in conjunction with the quantitative match between the measured and calculated quasiparticle bandgap, confirms the tip-induced electronic decoupling of N-6-ZGNRs from the underlying metallic Au(111) substrate.

*Ab initio* calculations on N-6-ZGNR show that an antiferromagnetic alignment of spins (across the ribbon width) between ferromagnetically-ordered edge states (e.g., as shown in Fig. 4c — left edge ↑↑, right edge ↓↓; the absolute spin orientation direction is arbitrary due to negligibly small spin-orbit interactions) is favoured as the ground state over a spatially spin-unpolarized configuration. The spin-polarization energy is 16 meV per edge atom, indicating large magnetic interaction energies. The antiferromagnetic spin configuration in the ground state is consistent with Lieb's theorem for interacting electrons on a bipartite lattice[28]. The GW LDOS of up (↑) and down (↓) spin integrated over the left and right half of a N-6-ZGNR (Fig. 4b) shows the expected spatial distribution of the two spin species at the bottom CBs and top VBs. Besides the obvious polarization of the frontier bands, our calculations predict two low-lying highly spatially spin-polarized states at $E–E_F$ = –2.60 eV and $E–E_F$ = –2.72 eV, respectively. These highly localized flat-band states are formed by the lone-pair orbitals of trigonal planar N-dopant atoms lining the edges of N-6-ZGNRs. The two bands (which show no splitting in spin-unpolarized calculations) split into an upper nitrogen flat band (UNFB; –2.60 eV) and a lower nitrogen flat band (LNFB;

–2.72 eV) due to the exchange interaction of the lone-pair electron with the spin-polarized $\pi$-electrons, which have a net spin population of opposite sign along the two edges. The exchange field generated by the ferromagnetic ordering on either N-6-ZGNR edge (corresponding to an effective Zeeman field of $B_{calc.}$ ~ 1000 T) favours an antiferromagnetic combination of lone-pair orbitals on the two edges with a local spin orientation that is parallel to the local exchange field. The LNFB and UNFB with spatial spin polarization in the same (left edge ↑, right edge ↓) and opposite (left edge ↓, right edge ↑) direction as the magnetic ordering of the N-6-ZGNR edges are lowered and raised in energy, respectively, leading to a splitting of the flat N-lone-pair bands.

To validate our theoretical predictions, we recorded d$I$/d$V$ point spectra along the edge of a fully decoupled N-6-ZGNR at a bias from $V_s$ = –1.50 V to $V_s$ = –3.00 V (Fig. S7). Only when the STM tip is located immediately above the position of a N-atom (Fig. 5a), the d$I$/d$V$ point spectra show two characteristic peaks centred at $V_s$ = –2.60 ± 0.02 V and $V_s$ = –2.70 ± 0.02 V, corresponding to the UNFB and LNFB states, respectively. d$I$/d$V$ imaging of the spatial distribution of the LDOS at energies close to UNFB and LNFB energies reveals a featureless ZGNR backbone that only shows enhanced contrast at the edge of the ribbon at the precise position of the N-dopant atoms (Figs. 5b,d, Fig. S8). The distinctive patterns in the d$I$/d$V$ maps faithfully reproduce the calculated UNFB and LNFB LDOS maps (Figs. 5c,e) further corroborating the assignment of UNFB and LNFB to the spin split N-lone-pair bands.

While the longitudinal modulation imposed by a substitutional N-dopant superlattice represents a universal strategy to stabilize and decouple the exotic magnetic edge states in ZGNRs, the isoelectronic substitution of a sixth of the C–H groups along the zigzag edge with N-atoms does not deteriorate the spin polarization and the edge magnetism in ZGNRs. Our calculated spatial distribution of electron spin polarization in the ground state (Fig. 4c) reveals that the magnetization on one N-atom in N-6-ZGNRs amounts to ~80% of the expected magnetization of a C-atom along the edge of a pristine 6-ZGNR, indicating almost unaltered magnetic edge states. The experimentally observed large exchange splitting $\Delta E_{(STS)}$ = 100 ± 30 meV of the low-lying N-lone-pair dopant flat bands by the ferromagnetically ordered spins along the edge of N-6-ZGNRs implies that the two electrons occupying the N-lone-pair experience an effective local exchange field $B_{eff} = 850 \pm 250$ T, consistent with the theoretical prediction. We therefore conclude that the isoelectronic substitution of C–H groups with N-atom dopants does not disrupt the intrinsic magnetization emerging from the spin-polarized edge states in ZGNRs.

Our results provide a general strategy to passivate the chemically reactive edges of zigzag nanographene without sacrificing the emergent spin degree of freedom and provide experimental evidence for the antiferromagnetic coupling of the ferromagnetically ordered edge states on opposite edges of ZGNRs. This approach creates a path for the development of atomically precise graphene-based high-speed low-power spin-logic devices for data storage and information processing.

**Acknowledgements**

Research supported by the Office of Naval Research under Award No. N00014-19-1-2503 (STM characterization), by the US Department of Energy (DOE), Office of Science, Basic Energy Sciences (BES) under the Nanomachine Program award number DE-AC02-05CH11231 (surface growth, image analysis), the National Science Foundation under Grant No. DMR-1839098 (molecular design and synthesis), Grant No. DMR-1926004 (GW calculations), and the Center for Energy Efficient Electronics Science ECCS-0939514 (theoretical analysis), the Office of Naval Research MURI under Award No. N00014-16-1-2921 (*ab initio* DFT calculations). Computational resources were provided by the DOE at Lawrence Berkeley National Laboratory NERSC facility and the NSF through XSEDE resources at NICS. R.E.B. acknowledges support through a National Science Foundation Graduate Research Fellowship under grant DGE-11064000.

**Author Contributions**

R.E.B., F.Z., S.G.L, and F.R.F. initiated and conceived the research, E.B., I.P. and F.R.F designed, synthesized, and characterized the molecular precursors, R.E.B., S.W., J.Z., A.D., and F.R.F. performed on-surface synthesis and STM characterization and analysis, F.Z., Y.-L.L., and S.G.L. performed DFT and GW calculations as well as theoretical analyses and assisted with data interpretation. R.E.B., F.Z., S.G.L. and F.R.F. wrote the manuscript. All authors contributed to the scientific discussion.

**Author Information**

The authors declare no competing financial interests. Correspondence and requests for materials should be addressed to ffischer@berkeley.edu or sglouie@berkeley.edu.

**Figure Legends**

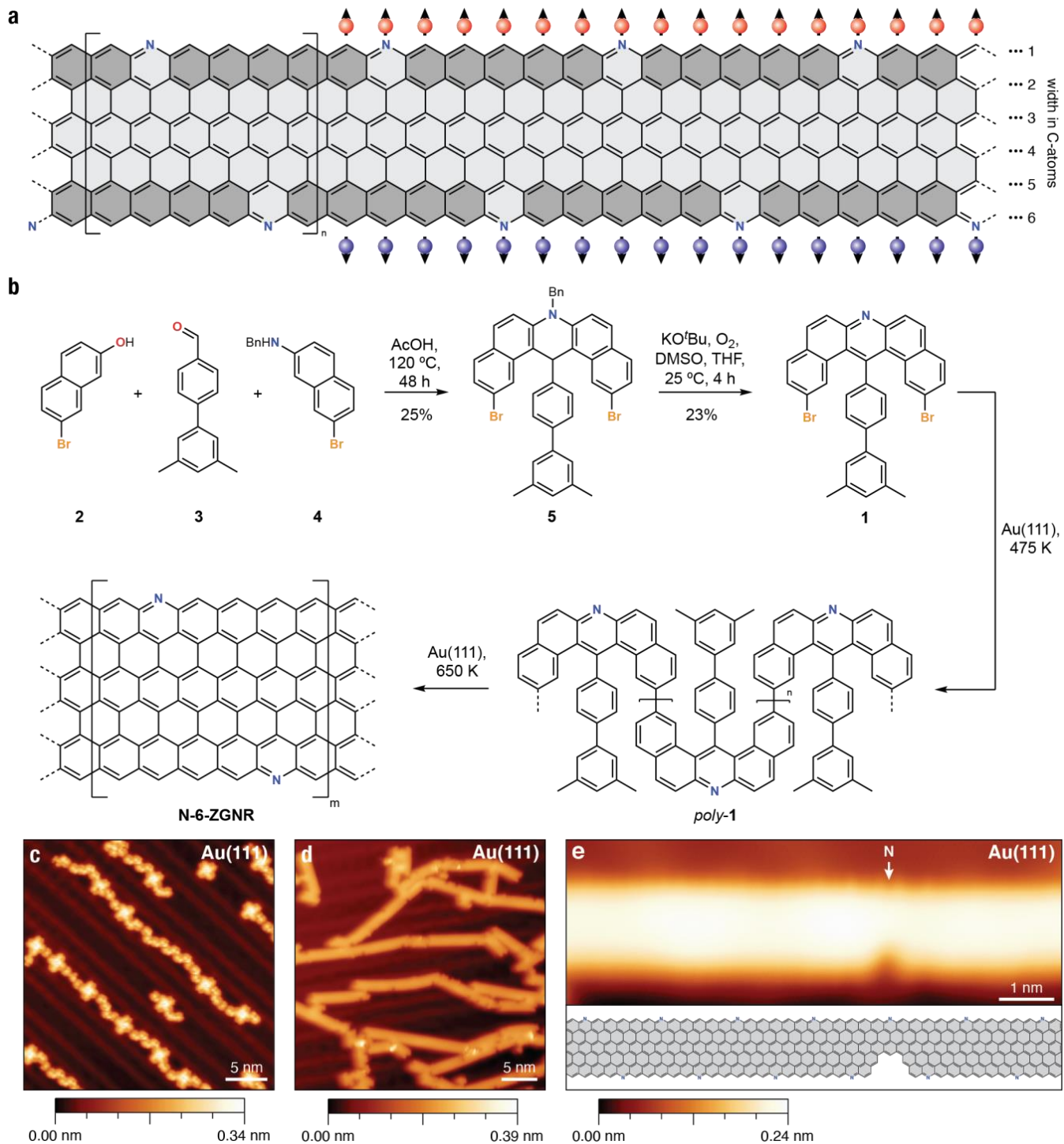

**Figure 1.** Bottom-up synthesis of N-doped N-6-ZGNRs. **a,** Schematic representation of spin-ordered edge states in N-6-ZGNRs. N-dopant superlattice modulates the spin-ordered edge states to segments reminiscent of pentacene units (highlighted in gray). **b,** Schematic representation of the bottom-up synthesis and on-surface growth of N-6-ZGNRs from molecular precursor **1**. **c,** STM topographic image of molecular precursor **1** as deposited on Au(111) ($V_s$ = 50 mV, $I_t$ = 20 pA). **d,** STM topographic image of fully cyclized N-6-ZGNRs following annealing to 650 K ($V_s$ = 50 mV, $I_t$ = 20 pA). **e,** STM topographic image of a fully cyclized N-6-ZGNR showing characteristic single point defect resulting from the cleavage of an *m*-xylene group ($V_s$ = 50 mV, $I_t$ = 20 pA). Arrow marks the position of a N-atom. Inset, chemical structure of the N-6-ZGNR segment imaged in (e). All STM images recorded at $T$ = 4 K.

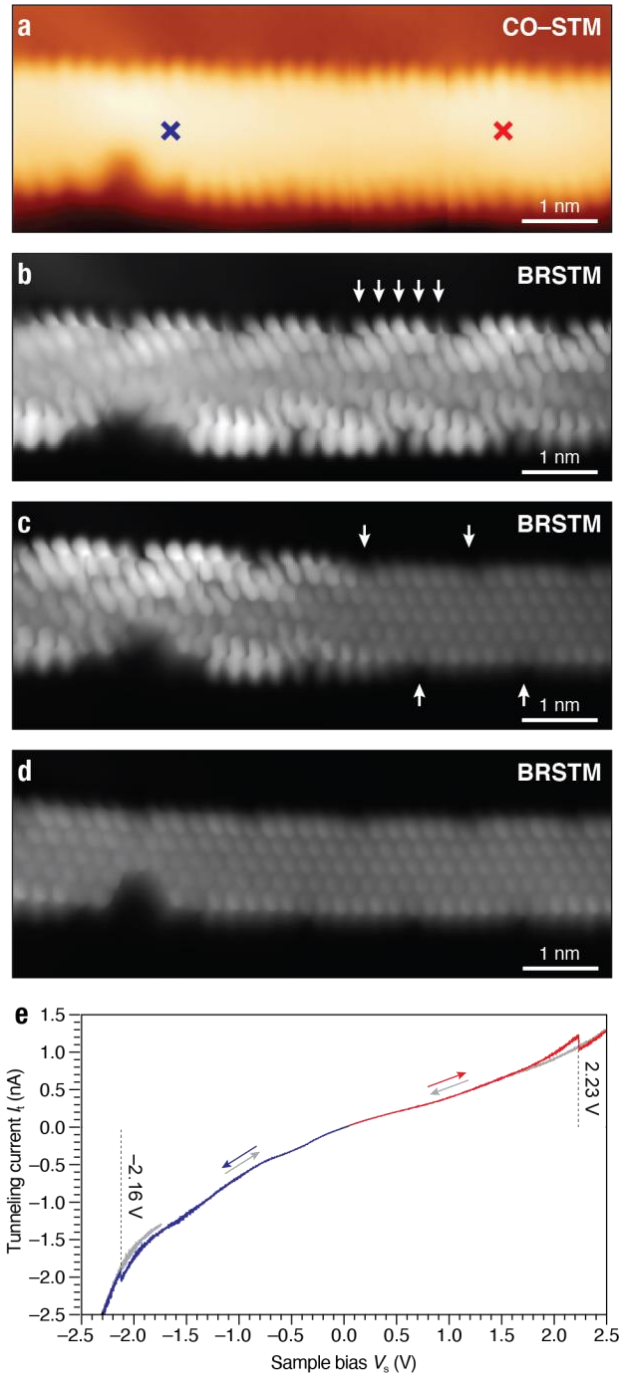

**Figure 2.** Tip-induced decoupling of magnetic edge states in N-6-ZGNRs from Au surface. **a,** Topographic image of a fully cyclized N-6-ZGNR segment recorded with CO-functionalized STM tip. **b,** Constant-height BRSTM image of the N-6-ZGNR segment from (a). Arrows mark the position of the five lobes associated with the C–H groups along the edges of the N-6-ZGNR. ($V_s$ = 0 mV, modulation voltage $V_{ac}$ = 11 mV, modulation frequency $f$ = 455 Hz). **c,** Constant-height BRSTM image of N-6-ZGNR segment

following tip-induced decoupling using a positive voltage sweep from $V_s = 0.0$ V to $V_s = +2.5$ V at the position marked by a red cross in (a) ($V_s = 0$ mV, $V_{ac} = 11$ mV, $f = 455$ Hz). Arrows mark the position of selected N-atoms along the edge of the N-6-ZGNR. **d,** BRSTM image of N-6-ZGNR segment following tip-induced decoupling using a negative voltage sweep from $V_s = 0.0$ V to $V_s = -2.5$ V at the position marked by a blue cross in (a) following decoupling in (c) ($V_s = 0$ mV, $V_{ac} = 11$ mV, $f = 455$ Hz). **e,** $I_t/V_s$ plot showing the positive (red) and negative (blue) voltage sweeps used during the decoupling procedure. Return sweeps (gray) show the irreversible shift in the tunneling current $I_t$. All STM experiments performed at $T = 4$ K

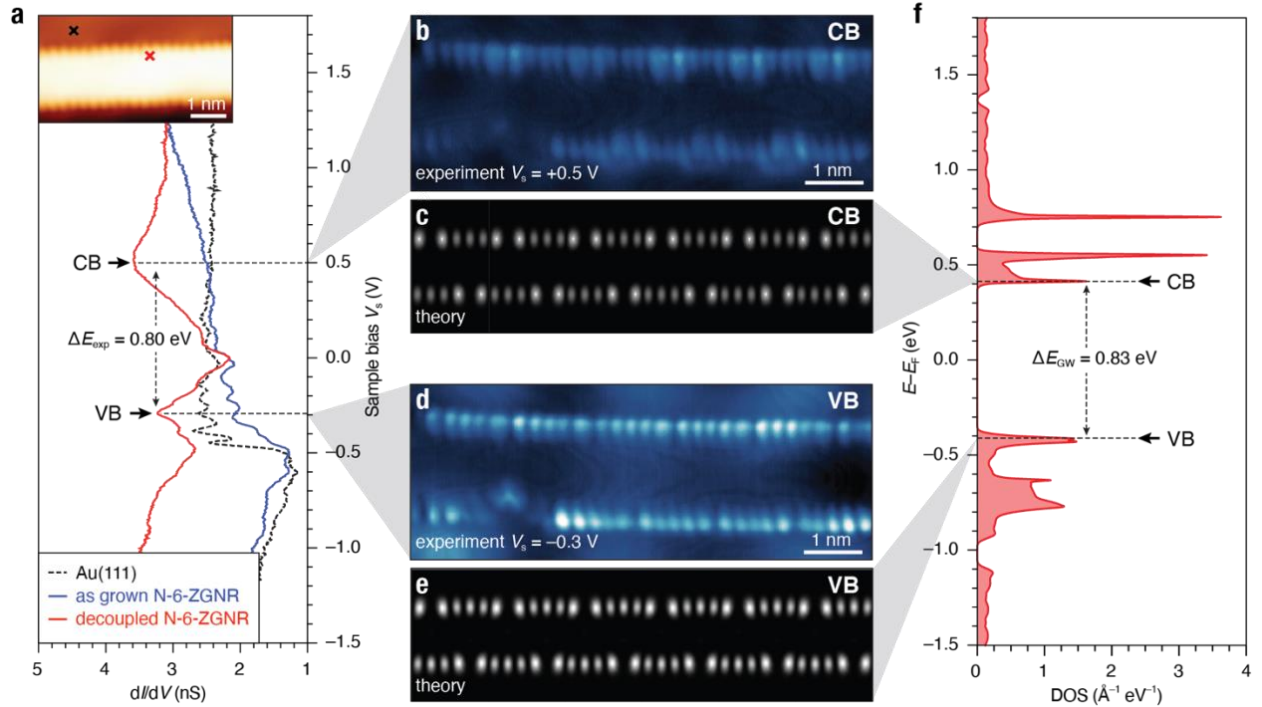

**Figure 3.** Electronic structure of N-6-ZGNR. **a,** d$I$/d$V$ point spectroscopy of as grown (blue) and decoupled (red) N-6-ZGNR/Au(111) at the position marked in the inset (red cross, imaging: $V_s$ = 50 mV, $I_t$ = 20 pA, CO functionalized tip); Au(111) reference spectrum (black) (spectroscopy: $V_{ac}$ = 11 mV, $f$ = 455 Hz). **b,** Constant-current d$I$/d$V$ map recorded at a voltage bias of $V_s$ = +0.5 V ($V_{ac}$ = 11 mV, $I_t$ = 200 pA, $f$ = 455 Hz, CO functionalized tip). **c,** Calculated GW LDOS at the conduction band (CB) edge. **d,** Constant-current d$I$/d$V$ map recorded at a voltage bias of $V_s$ = –0.3 V ($V_{ac}$ = 11 mV, $I_t$ = 170 pA, $f$ = 455 Hz, CO functionalized tip). **e,** Calculated GW LDOS at the valence band (VB) edge. **f,** Calculated GW DOS for a N-6-ZGNR (spectrum broadened by 4 meV Gaussian). All STM experiments were performed at $T$ = 4 K. Theoretical LDOS are sampled at a height of 4 Å above the atomic plane of the N-6-ZGNR.

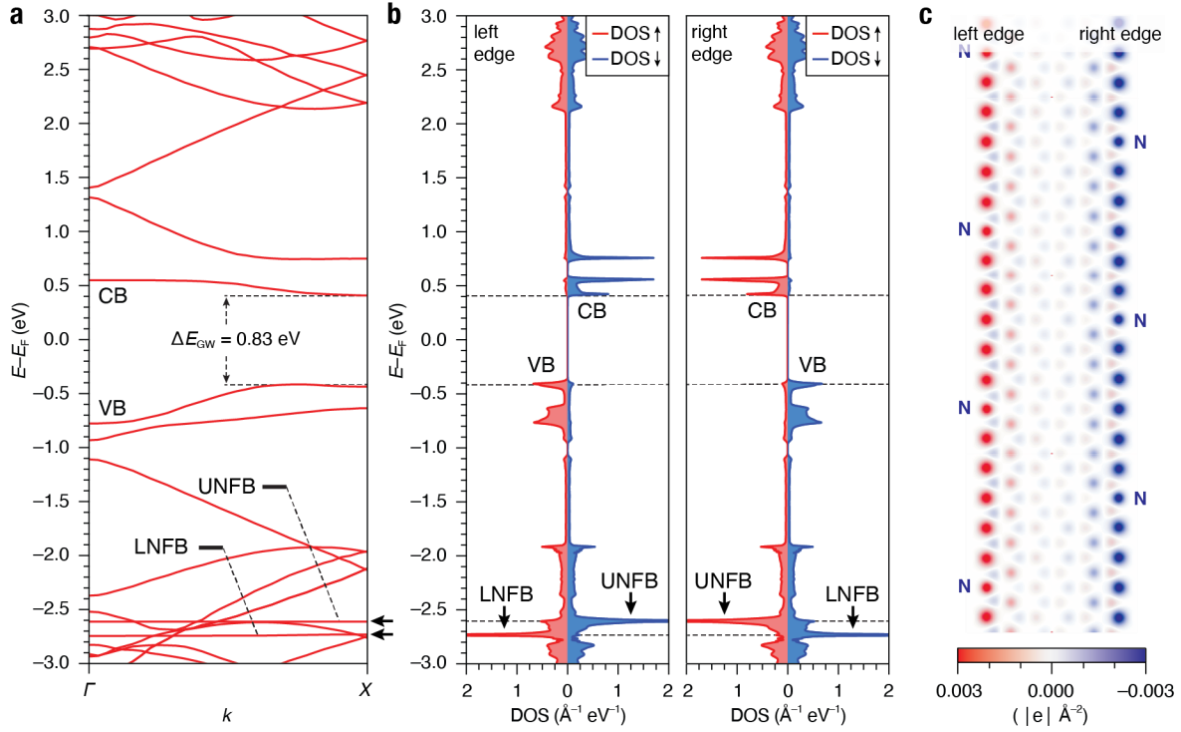

**Figure 4.** Band structure and spatial distribution of spin-ordered edge states in N-6-ZGNRs. **a,** GW band structure of a freestanding N-6-ZGNR. Upper (UNFB) and lower (LNFB) nitrogen flat bands are highlighted by arrows. **b,** GW LDOS of up (red) and down (blue) spin integrated over the left half and right half of a N-6-ZGNR as shown in (C). **c,** Spatial distribution of the areal spin density distribution difference between up and down spin $(\rho_\uparrow(r) - \rho_\downarrow(r))$. The areal density is the density integrated in the direction out of the GNR atomic plane.

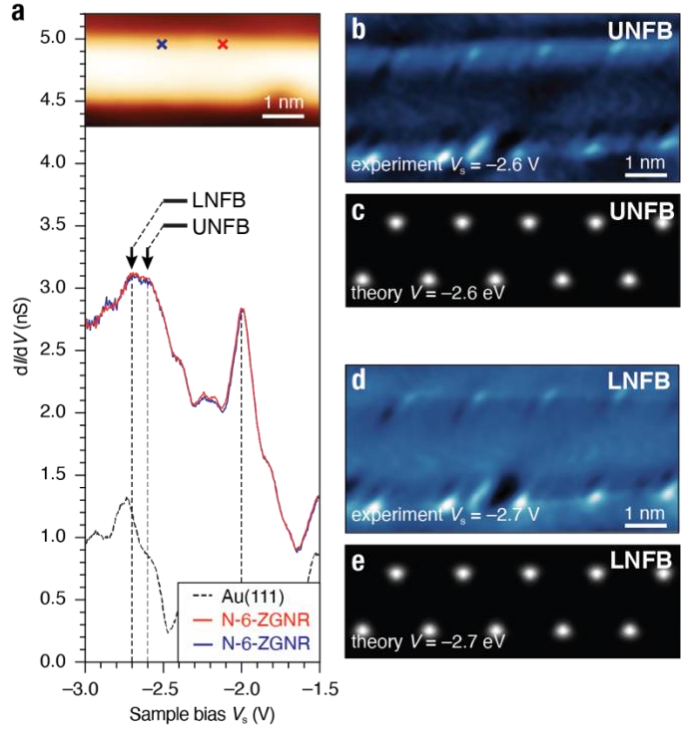

**Figure 5.** Spin splitting of nitrogen flat band states (NFB) of $sp^2$ lone-pair orbitals in N-6-ZGNRs. **a,** d$I$/d$V$ point spectroscopy of decoupled N-6-ZGNR/Au(111) at the position of two N-atoms marked in the inset (red and blue cross, imaging: $V_s$ = 300 mV, $I_t$ = 20 pA); Au(111) reference spectrum (black) (spectroscopy: $V_{ac}$ = 11 mV, $f$ = 455 Hz). **b,** Constant-current d$I$/d$V$ map of UNFB recorded at a voltage bias of $V_s$ = –2.6 V ($V_{ac}$ = 11 mV, $I_t$ = 20 pA, $f$ = 455 Hz). **c,** Calculated GW LDOS of UNFB. **d,** Constant-current d$I$/d$V$ map of LNFB recorded at a voltage bias of $V_s$ = –2.7 V ($V_{ac}$ = 11 mV, $I_t$ = 20 pA, $f$ = 455 Hz). **e,** Calculated GW LDOS of LNFB.

**Methods**

**Precursor Synthesis and GNR Growth.** Full details of the synthesis and characterization of **1**–**5** are given in the Supplementary Information. N-6-ZGNRs were grown on Au(111)/mica films under UHV conditions. Atomically clean Au(111) surfaces were prepared through iterative Ar$^+$ sputter/anneal cycles. Sub-monolayer coverage of **1** on atomically clean Au(111) was obtained by sublimation at crucible temperatures of 460–470 K using a home-built Knudsen cell evaporator. After deposition the surface temperature was slowly ramped ($\leq$ 2 K min$^{-1}$) to 475 K and held at this temperature for 30 min to induce the radical-step growth polymerization, then ramped slowly ($\leq$ 2 K min$^{-1}$) to 650 K and held there for 30 min to induce cyclodehydrogenation.

**STM Measurements.** All STM experiments were performed using a commercial OMICRON LT-STM held at $T$ = 4 K using PtIr STM tips. STM tips were optimized for scanning tunneling spectroscopy using an automated tip conditioning program[29]. d$I$/d$V$ measurements were recorded with CO-functionalized STM tips using a lock-in amplifier with a modulation frequency of 455 Hz and a modulation amplitude of $V_\text{RMS}$ = 11 mV. d$I$/d$V$ point spectra were recorded under open feedback loop conditions. d$I$/d$V$ maps were collected under constant current conditions. BRSTM images were obtained by mapping the out-of-phase d$I$/d$V$ signal collected during a constant-height d$I$/d$V$ map at zero bias. Peak positions in d$I$/d$V$ point spectroscopy were determined by fitting the spectra with Lorentzian peaks. Each peak position is based on an average of approximately 90 spectra collected on 10 GNRs with 10 different tips, all of which were first calibrated to the Au(111) Shockley surface state.

**Tip Induced Decoupling Protocol.** From the tunneling setpoint ($I_\text{t}$ = 20 pA, $V_\text{s}$ = 50 mV), the STM tip is relocated to a position above the Au(111) substrate. Using an open-feedback loop, the bias voltage is ramped to ± 2.5 V at a rate of approximately 100 mV s$^{-1}$ and the tunneling current as a function of time is monitored. A monotonically increasing (decreasing) $I_\text{t}$ curve is used to ensure that a global tip change has not occurred. The process is repeated with the STM tip located above the GNR backbone. Discontinuous drops in the tunneling current are used to indicate a successful decoupling event.

**Calculations.** DFT calculations of GNR superlattices were performed in the LSDA as implemented in the Quantum ESPRESSO package[30], and the GW calculation is performed by the BerkeleyGW package[31]. Only freestanding GNRs (without substrate) were calculated. A supercell arrangement was used with vacuum regions carefully tested to avoid interactions between the superlattice and its periodic images. We used norm-conserving pseudopotentials with a planewave energy cut-off of 60 Ry. The structure was fully relaxed within DFT-LSDA until the magnitude of the force on each atom was smaller than 0.02 eV Å$^{-1}$. All σ dangling bonds on the edges and the ends of the GNRs were capped by hydrogen atoms. A Gaussian broadening of 4 meV was used in the evaluation of the DOS and LDOS. In the *GW* calculation, the frequency-dependent screening is incorporated by the Hybertson-Louie GPP model[27], and the self-consistency in the quasiparticle energy eigenvalues in the Green's function *G* is reached.

**Data Availability**

DFT code with pseudopotentials and GW code can be downloaded from the Quantum Espresso[32] and the BerkeleyGW[33] website, respectively. For this study we used Quantum Espresso version 6.4.1 and BerkeleyGW version 2.1 for the theoretical calculations. All data presented in the main text and the supplementary materials are available from the corresponding authors upon reasonable request.

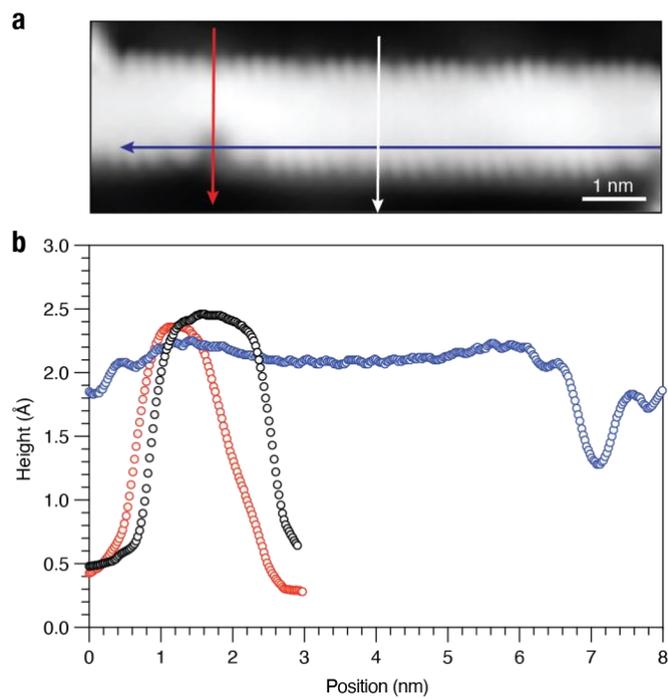

**Figure S1.** Height profiles of N-6-ZGNRs on Au(111). **a,** STM topographic image of N-6-ZGNR on Au(111) ($V_s$ = 50 mV, $I_t$ = 20 pA; CO-functionalized tip). **b,** Height profile recorded along the arrows marked in (a).

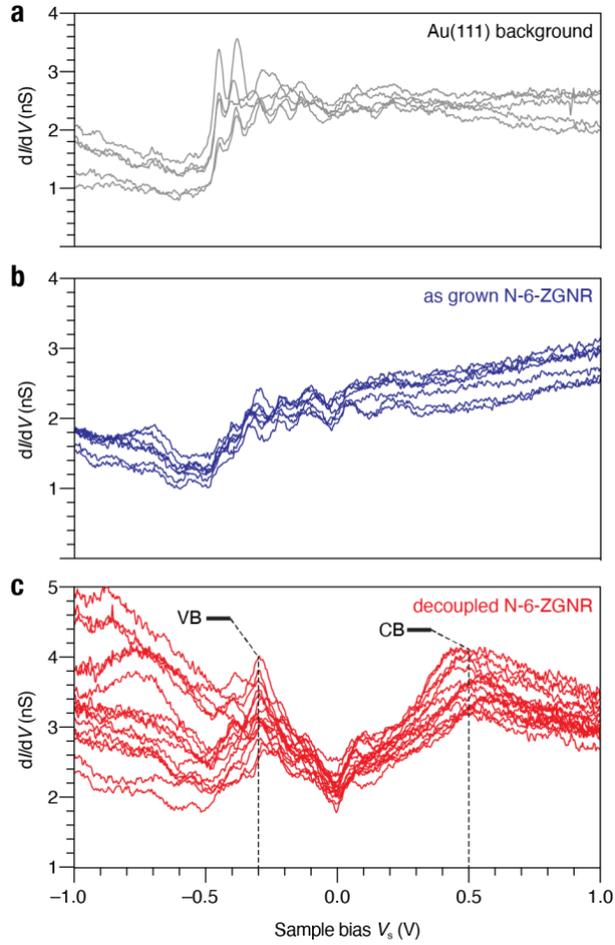

**Figure S2.** d$I$/d$V$ point spectroscopy of N-6-ZGNRs on Au(111). **a,** d$I$/d$V$ point spectra collected on bare Au(111) ($V_{ac}$ = 11 mV, $f$ = 455 Hz). **b,** d$I$/d$V$ point spectra collected on as grown segments of N-6-ZGNRs. **c,** d$I$/d$V$ point spectra collected on decoupled segments of N-6-ZGNRs following the SPM tip-activation protocol.

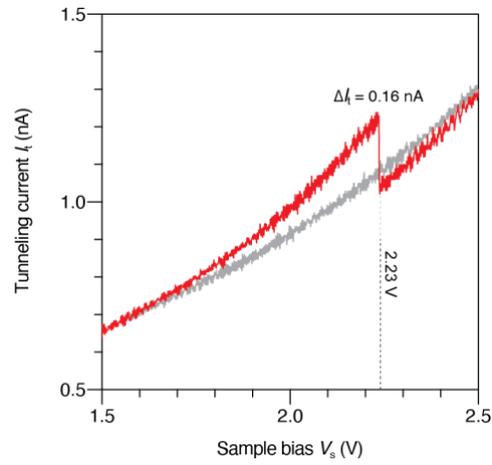

**Figure S3.** Tip induced decoupling of N-6-ZGNRs. $I_t/V_s$ plot showing in red the positive voltage sweep ($V$s = +1.50 V to $V$s = +2.50 V) used during the decoupling procedure. The respective return sweep ($V$s = +2.50 V to $V$s = +1.50 V) is depicted in gray and shows the irreversible shift in the tunneling current $I_t$.

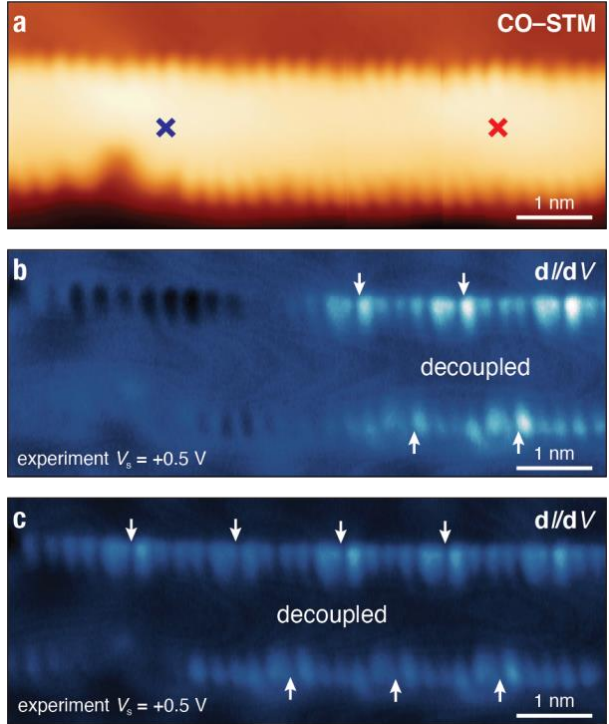

**Figure S4.** Tip-induced decoupling of magnetic edge states in N-6-ZGNRs. **a,** Topographic image of a fully cyclized N-6-ZGNR segment recorded with CO-functionalized STM tip. **b,** Constant-current d$I$/d$V$ map recorded at a voltage bias of $V_s$ = +0.5 V of N-6-ZGNR segment following tip-induced decoupling using a positive voltage sweep from $V_s$ = 0.0 V to $V_s$ = +2.5 V at the position marked by a red cross in (a) ($V_{ac}$ = 11 mV, $I_t$ = 200 pA, $f$ = 455 Hz, CO functionalized tip). **c,** Constant-current d$I$/d$V$ map recorded at a voltage bias of $V_s$ = +0.5 V of N-6-ZGNR segment following tip-induced decoupling using a negative voltage sweep from $V_s$ = 0.0 V to $V_s$ = –2.5 V at the position marked by a blue cross in (a) ($V_s$ = 0 mV, $V_{ac}$ = 11 mV, $f$ = 455 Hz). Arrows mark the position of selected N-atom along the edge of the N-6-ZGNR.

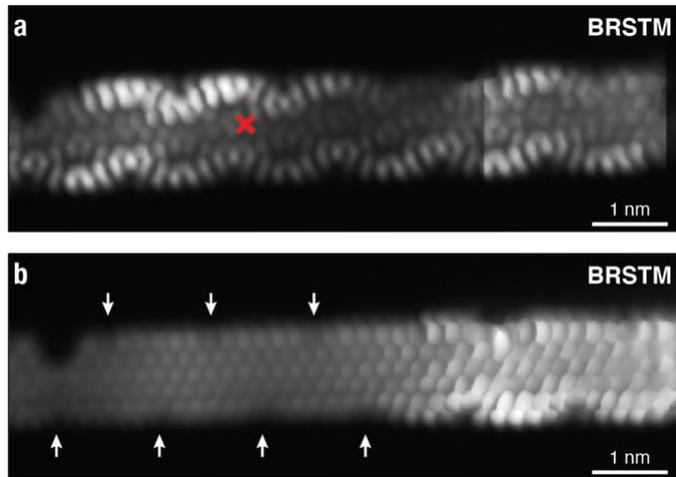

**Figure S5.** Tip-induced decoupling of magnetic edge states in N-6-ZGNRs. **a,** Constant-height BRSTM image of a N-6-ZGNR segment from ($V_s$ = 0 mV, modulation voltage $V_{ac}$ = 11 mV, modulation frequency $f$ = 455 Hz). **b,** Constant-height BRSTM image of the N-6-ZGNR segment in (a) following tip-induced decoupling using a positive voltage sweep from $V_s$ = 0.0 V to $V_s$ = +2.5 V at the position marked by a red cross in (a) ($V_s$ = 0 mV, $V_{ac}$ = 11 mV, $f$ = 455 Hz). Arrows mark the position of selected N-atom along the edge of the N-6-ZGNR.

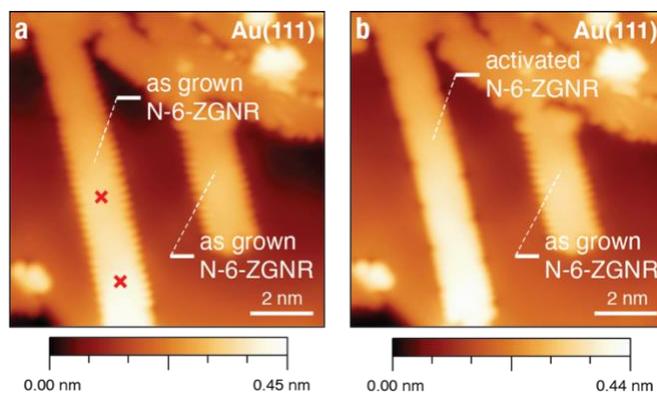

**Figure S6.** STM images of partially decoupled N-6-ZGNRs. **a,** STM topographic image of as-grown N-6-ZGNRs with CO-modified tip. **b,** STM topographic image of N-6-ZGNR after decoupling the GNR on the left. ($V_s$ = 50 mV, $I_t$ = 20 pA)

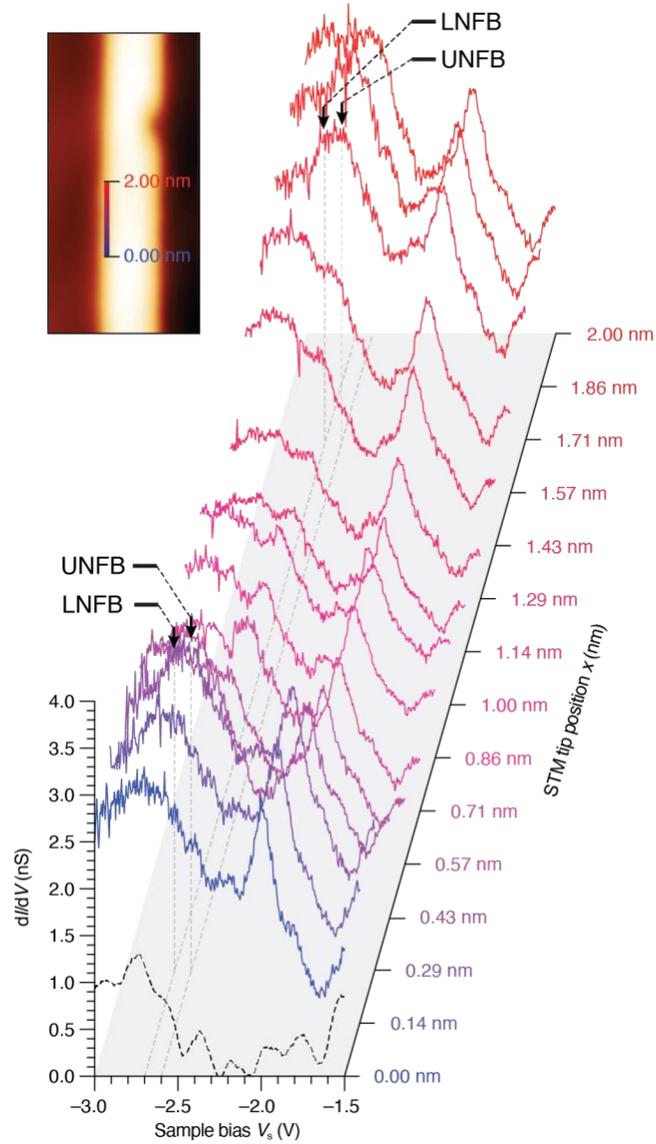

**Figure S7.** d$I$/d$V$ point spectroscopy of spin split low-lying nitrogen dopant flat band states. Vertically stacked d$I$/d$V$ point spectra collected along a line marked in the inset long the edge of a N-6-ZGNR ($V_{ac}$ = 11 mV, $f$ = 455 Hz). When the STM tip is located immediately above the position of a nitrogen dopant atom ($x$ = 0.29 nm, and $x$ = 1.71 nm) the d$I$/d$V$ point spectra show two characteristic peaks centred at $V_s$ = −2.60 ± 0.02 V and $V_s$ = −2.70 ± 0.02 V, corresponding to the UNFB and LNFB states, respectively.

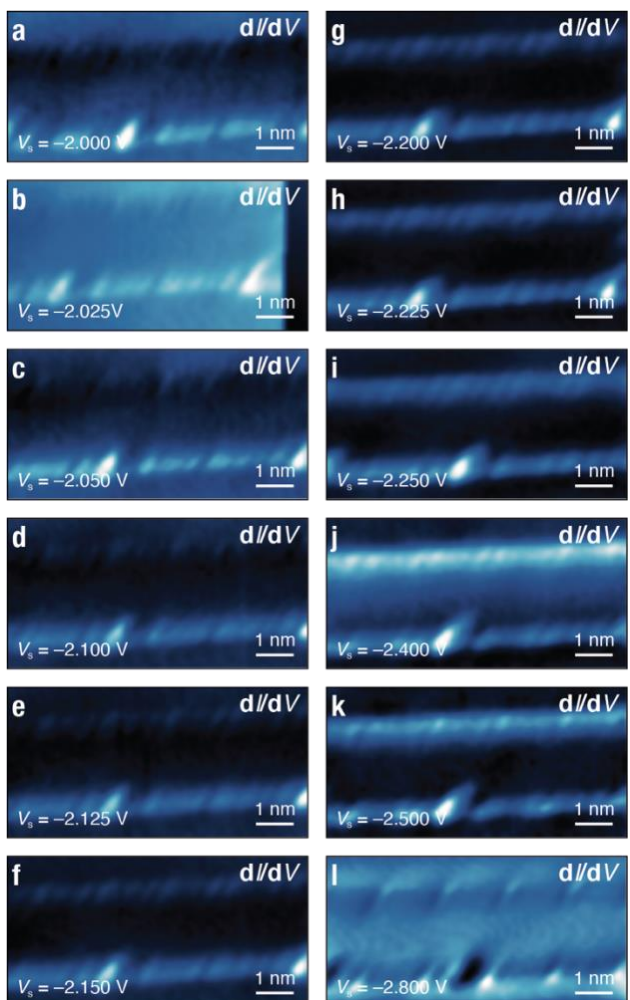

**Figure S8.** Constant-current d$I$/d$V$ maps of decoupled N-6-ZGNR. d$I$/d$V$ maps recorded at voltage biases of **a,** $V_s = -2.000$ V, **b,** $V_s = -2.025$ V, **c,** $V_s = -2.050$ V, **d,** $V_s = -2.100$ V, **e,** $V_s = -2.125$ V, **f,** $V_s = -2.150$ V, **g,** $V_s = -2.200$ V, **h,** $V_s = -2.250$ V, **i,** $V_s = -2.050$ V, **j,** $V_s = -2.400$ V, **k,** $V_s = -2.500$ V, and **l,** $V_s = -2.800$ V ($V_{ac} = 11$ mV, $I_t = 200$ pA, $f = 455$ Hz).

**Supplementary Information**

Supplementary Information contains detailed synthesis and characterization of precursor **1.**